\documentclass[
aps,
pra,
twocolumn,
groupedaddress,
numbers,
nofootinbib]{revtex4-1}
\usepackage{times}
\usepackage[T1]{fontenc} 
\usepackage[utf8]{inputenc} 
\usepackage[english]{babel} 
\usepackage{graphicx, xcolor}
\usepackage{amsfonts, amssymb, amsmath, mathrsfs, calc, array, mathtools}
\usepackage[separate-uncertainty=true]{siunitx}
\usepackage{hyperref}
\usepackage{abraces}
\usepackage{empheq}

\newcommand*\widefbox[1]{\fbox{\hspace{0.em}#1\hspace{0.em}}}

\def\bra#1{\mathinner{\langle{#1}|}}
\def\ket#1{\mathinner{|{#1}\rangle}}

\def\ontop#1#2{\setbox0\hbox{#2}\copy0\llap{\raise\ht0\hbox{#1}}}

\newcommand{\defeq}{\vcentcolon=}

\newcommand{\ic}{\mathrm{i}}

\definecolor{darkblue}{rgb}{0,0,0.93} 
\definecolor{darkred}{rgb}{0.8,0,0} 

\hypersetup{
colorlinks=true, 
linkcolor=darkblue, 
citecolor=darkred, 
filecolor=darkblue, 
urlcolor=darkblue
}

\newcommand{\T}{\mathcal{T}_1}


\begin{document}


\title{Clifford algebra from quantum automata and unitary Wilson fermions}


\author{Pablo Arnault}
\email{pablo.arnault@inria.fr}
\affiliation{Laboratoire Méthodes Formelles, Université Paris-Saclay, CNRS, ENS Paris-Saclay, CentraleSupélec, INRIA, F-91190 Essonne, France}



\begin{abstract}
We introduce a spacetime discretization of the Dirac equation that has the form of a quantum automaton and that is invariant upon changing the representation of the Clifford algebra, as the Dirac equation itself. Our derivation follows Dirac's original one: We required that the square of the discrete Dirac scheme be what we define as an acceptable discretization of the Klein-Gordon equation. Contrary to standard lattice gauge theory in discrete time, in which unitarity needs to be proven, we show that the quantum automaton delivers naturally unitary Wilson fermions for any choice of Wilson's parameter.
\end{abstract}


\maketitle


\vspace{-1.8cm}

\emph{{\bfseries Introduction.}}
Symmetries are at the grounding of physics.
Indeed, when trying to establish a law empirically, one typically starts by discarding irrelevant parameters by identifying the symmetries of the problem under study, e.g., spatial or temporal symmetries.
To actually establish a symmetry empirically, one has to reproduce an experiment several times, making physically vary the parameter suspected not to be relevant in the law, e.g., changing the location of the experiment or the time at which it is carried out, and see if the result of the experiment is still the same.
Thus, a given physical law has the mathematical invariances, also called symmetries, that reflect the physical symmetries of the phenomena it describes.

The Dirac equation (DE) is the law of motion of relativistic quantum particles of matter (more precisely, of so-called fermionic particles), and is one of the main equations of the standard model of particle physics.
One fundamental property of the DE is that its square delivers the Klein-Gordon equation.
Now, the constraints imposed by this property on the equation take the form of algebraic relationships, the so-called Clifford algebra, between objects that enter the definition of the DE, the so-called gamma matrices.
Any particular family of gamma matrices that satisfy the Clifford algebra is thus a good candidate to write down the DE and describe with it the appropriate physics; such a family is called a representation of the Clifford algebra.
Hence, the DE has the following symmetry: it is invariant upon changing the Clifford-algebra representation.
This symmetry is fundamental: the physics is captured by the Clifford algebra, and not by a single, particular family of gamma matrices chosen.
Moreover, the Clifford algebra is at the grounding of modern geometric algebra \cite{LS21}, which shows its fundamental character beyond the DE, for understanding the geometry of our world.

It is well-known that numerically simulating real-time dynamics of quantum multiparticle systems is exponentially hard, and that quantum computers could overcome this difficulty.
The DE is a central ingredient of these quantum multiparticle systems in the relativistic regime.
In order to perform simulations involving the DE, one starts by discretizing it on a spacetime grid \cite{Wilson74, KogSuss75a}.
But, if one does so without care, one breaks certain symmetries satisfied by the DE.
Now, discrete-time quantum walks (DQWs) are quantum transport schemes in discrete spacetime that have been the subject of much attention in the last 30 years, in particular because of their success as discretizations of the DE that preserve various of its symmetries \cite{BB94a, Meyer96a, Bisio2015}, also when coupled to the fundamental force fields of nature \cite{AD16a, AD16b, ced13, DMD13b,  AD17, Arrighi_curved_1D_15, AF17, AMBD16}.
There are two main such symmetries that DQWs preserve: the first is unitarity;
the second, which is actually rather a property, is relativistic locality, i.e., the group velocity in vacuum cannot overcome the speed of light.
We call Dirac DQW a DQW corresponding to a discretization of the DE that exhibits the two aforementioned properties.

Up to our knowledge, no existing Dirac DQW has yet managed to preserve the invariance of the DE upon changing the Clifford-algebra representation.
More precisely, the representation chosen when discretizing the DE with a Dirac DQW is actually always the same, e.g., in $1+1$ dimensions, the first alpha matrix always equals the third Pauli matrix.
If one choses initially a different representation, one will need a different discretization method.
So, not only the structure of the resulting discretization, but also the method used to discretize the equation, depend on the Clifford-algebra representation, which is strongly unsatisfying.

In the present work, we give a solution to this issue, by introducing a Dirac-DQW-based discretization method of the DE that can be carried out whatever the Clifford-algebra representation.
More precisely, we make it possible the emergence of the Clifford algebra out of the operators defining the Dirac DQW.
This is achieved by choosing appropriate such operators, but still independent of any choice of basis; they only need to satisfy the above-mentioned Clifford algebra.
This Clifford algebra is obtained, following Dirac's own procedure in continuous spacetime, by requiring that the Dirac DQW squares a spacetime-discretized version of the Klein-Gordon equation that we consider valid, i.e., it must not only deliver the correct Klein-Gordon equation in the continuum limit, but also (i) be applicable to scalar state sequences \emph{at the discrete level}, as well as satisfying an extra condition, namely, (ii) the vanishing of the crossed term \emph{at the discrete level}.
We then show that DQWs also contain, ``naturally'' (provided we make certain appropriate choices), a Wilson term that makes fermion doubling be avoided, which is well-known in lattice gauge theory (LGT).
Unitarity is maintained at each step of the derivation, and this \emph{for any choice $r \in \mathbb{R}$ of Wilson's parameter}.

\emph{{\bfseries Discrete-time quantum walks.}} A DQW is a unitary automaton with an ultralocal evolution operator.
The system is called a \emph{walker}, and its state at time $j\in\mathbb{N}$ is given by a sequence $\Psi_j : (j,p) \mapsto \Psi_{j,p}$ defined over a lattice with sites labeled by $p \in \mathbb{Z}$; the lattice is considered $1\text{D}$ for simplicity. 
The dynamics of this system is called a \emph{walk}, written $\Psi_{j+1} = \mathcal{U} \Psi_j$, where $\mathcal{U}$ is the unitary one-step evolution \emph{walk operator}.
That $\mathcal{U}$ is \emph{ultralocal} means, by definition, that the internal state $\Psi_{j+1,p}$ solely depends on internal states $\Psi_{j,p'}$ that belong to a bounded spatial-lattice neighborhood around~$p$.
In the context of automata, ultralocality of $\mathcal{U}$ is often implicit.
In quantum computation it is frequent that $\mathcal{U}$ is not ultralocal: this can be either because we purposely apply, in discrete time, a gate which is non-local \cite{NHA20}, or because we are in continuous time and evolve a system via a nearest-neighbors Hamiltonian, which yields approximate but not exact ultralocality via Lieb-Robinson bounds \cite{LR72}.

To specify the nature of $\Psi_j$, it is useful to invoke Meyer's no-go result \cite{Meyer96a, Meyer96b}: No non-trivial ultralocal unitary automaton with a one-step evolution operator that is homogeneous in space (i.e., translationally invariant) can have a scalar walker; the minimum number of internal components for $\Psi_{j,p}$ is thus $2$.
Hence, we consider $\Psi_{j,p} \in \mathscr{H}_2$, a complex Hilbert space of dimension $2$.
In formal parallel with classical random walks \cite{Kempe2003a}, in which a coin is tossed to determine the direction taken by the walker, the internal state $\Psi_{j,p}$ is called \emph{coin state}, and $\mathscr{H}_2$ is called \emph{coin space}.
%

\emph{{\bfseries Transport coin operators of a DQW.}} 
Consider the DQW defined initially, $\Psi_{j+1} = \mathcal{U} \Psi_j$.
The one-step evolution operator $\mathcal{U} $ can be given under a multiplicative \cite{Vogts09, CGW21} or an additive form \cite{Meyer96a, Bisio2015}.
We here give $\mathcal{U}$ under the generic additive form $\mathcal{U} \defeq W_{-1} \T^{-1} + W_1 \T + W_0$.
In this equation, (i) $\T$ is the translation operator by one lattice site in the direction of growing $p$s (that is, $(\mathcal{T}_{1}\Psi)_{j,p}\defeq \Psi_{j,p-1} $), and (ii) the $W_i$s, $i=-1,0,1$, are operators acting solely on the coin space, that we call \emph{jump coin operators}.
While one may view the $W_i$s as $2 \times 2$ complex matrices, viewing them abstractly is actually the purpose of the present work, and \emph{we will not introduce any basis of $\mathscr{H}_2$}.
The translation operator is by construction related to the momentum operator, $\mathcal{K}$, via $\T = e^{-\ic a \mathcal{K}}$, where $a \in \mathbb{R}^{+, \ast}$ will be identified further on with the spatial lattice spacing.
In the Supplemental Material, we derive the constraints that the unitarity of $\mathcal{U}$ imposes on the $W_i$s \cite{Meyer96a, Bisio2015}.

We define the following coin operators, that we call \emph{transport coin operators},
\begin{subequations}
\begin{align}
B &\defeq W_{1} - W_{-1} \\
V &\defeq W_{1} + W_{-1} \\
M &\defeq {\small \sum}_{i = -1,0,1} W_i = V + W_0 \, .
\end{align}
\end{subequations}
\noindent
In terms of these operators, $\mathcal{U}$ reads $\mathcal{U} = \tfrac{1}{2}(V-B)\T^{-1} + \tfrac{1}{2}(V+B)\T + M - V$.
In the Supplemental Material, we translate on the transport coin operators the constraints imposed on the jump coin operators by the unitarity of $\mathcal{U}$.

\emph{{\bfseries Local Hamiltonian of a DQW.}} 
From the one-step dynamics $\Psi_{j+1} = \mathcal{U} \Psi_j$, one can conceive a dynamics ${\ic}(\Psi_{j+1} - \Psi_{j-1})/2 = \mathcal{H} \Psi_j$ determined by the Hermitean operator $\mathcal{H} \defeq \frac{\ic}{2}( \mathcal{U} - \mathcal{U}^{\dag} )$, which is (ultra)local since $\mathcal{U}$ is ultralocal, and that we call \emph{local Hamiltonian of the DQW}.
This dynamics is ``two-step'', meaning that while the one-step dynamics takes as initial condition $\Psi_{j=0}$, the two-step one takes as initial condition both ${\Psi_{j=0}}$ and ${\Psi_{j=1}}$.
The two-step dynamics is equivalent to the one-step one provided that $\Psi_1 = \mathcal{U} \Psi_0$.
That the Hamiltonian $\mathcal{H}$ is (ultra)local is in contrast with the case of the well-known effective Hamiltonian of the DQW.
Note that both Hamiltonians are related by a proportionality constant in Fourier space \cite{APP20}.

The operator $\mathcal{H}$ can be written in terms of the transport coin operators as
\begin{equation}
\label{eq:H_final}
\mathcal{H} = \mathcal{H}'_{Q} \defeq A^1 ( - \ic \mathcal{D}_1 ) + \frac{r}{2} Q (- \mathcal{L}) + \epsilon m A^0 \, ,
\end{equation}
where $m \in \mathbb{R}^{+}$ and $r \in \mathbb{R}$ are two parameters that we force to appear (we also force $\epsilon$ to appear in $\epsilon m$).
In Eq.\ \eqref{eq:H_final}, we have introduced (i) the following operators acting on the position space solely,
$\mathcal{D}_1 \defeq \frac{1}{2} (\T^{-1} - \T)$ and $\mathcal{L} \defeq \T^{-1} + \T - 2$,
and the following coin operators made out of the transport coin operators, $
A^1 \defeq (B + B^{\dag})/2$, $Q \defeq - \tfrac{\ic}{r} (V - V^{\dag})/2$, and
$A^0 \defeq \tfrac{\ic}{\epsilon m} (M - M^{\dag})/2$.
We have used the notation $\mathcal{H} = \mathcal{H}'_Q$ because we are going to consider both the case $Q = 0$ and the case $Q \neq 0$.

\emph{ {\bfseries Dirac-continuum-limit requirement.}} 
The two-step dynamics can be written
\begin{equation}
\label{eq:two-step}
 \ic (\mathcal{D}_0 \Psi)_j = \mathcal{H} \Psi_j \, ,
\end{equation}
where $\mathcal{D}_0 \defeq ( \mathcal{T}_{0}^{-1} - \mathcal{T}_{0} )/2$, with $\mathcal{T}_{0}$ the shift by one lattice site forward in time, i.e., $
(\mathcal{T}_{0}\Psi)_j \defeq \Psi_{j-1} $.
We wish that the two-step dynamics deliver the $(1+1)\text{D}$ DE in the continuum limit.
In order to take the continuum limit, we introduce continuous time and space coordinates $t$ and $x$, as well as a function of these continuous coordinates, $\Psi(\cdot,\cdot) : (t,x) \mapsto \Psi(t,x)$, that is as smooth as wished, and that coincides at coordinates $(t_j \defeq j \epsilon, x_p \defeq p a)$ with the ``value'' taken by the coin state at point $(j,p)$, that is, $\Psi(t_j,x_p) \defeq  \Psi_{j,p}$.
We then consider the ballistic scaling $\epsilon = a$ \cite{DDMEF12a}, and Taylor expand the two-step dynamics in time and space around the point $(t_j,x_p)$, at second order in $\epsilon$.

For the continuum limit of the two-step dynamics, Eq.\ \eqref{eq:two-step} (divide Eq.\  \eqref{eq:two-step}  by $\epsilon$ and let $\epsilon \rightarrow 0$ after the Taylor expansion at second order), to coincide with the $(1+1)\text{D}$ DE, it is sufficient that the two following \emph{Dirac-continuum-limit constraints} be satisfied,
$A^1 \defeq \frac{B+B^{\dag}}{2} \underset{\epsilon \rightarrow 0}{\sim} \alpha^1\label{eq:Dirac_constraint1}$ and
$\epsilon m A^0 \defeq \ic \, \frac{M - M^{\dag}}{2} \underset{\epsilon \rightarrow 0}{\sim} \epsilon m \alpha^0$,
where $\alpha^1$ and $\alpha^0$ must satisfy $(\alpha^0)^2 = (\alpha^1)^2 = 1$ and $\alpha^0 \alpha^1 +\alpha^1 \alpha^0 = 0$, in order for them to correspond to the well-known $\alpha$ operators of the DE.

Note that no continuum-limit constraint is imposed on $Q$ and thus neither on $V$ (apart from $Q$ not scaling as $\epsilon^{\delta}$ with $\delta \leq - 1$, i.e., we must have $\delta > - 1 $), because as $\epsilon \rightarrow 0$, we have that $\mathcal{L} \sim \epsilon^2 \partial_x^2$ while $\mathcal{D}_1 \sim \epsilon \partial_x$, so that the term $\frac{r}{2} Q (- \mathcal{L})$ in Eq.\ \eqref{eq:H_final} vanishes in the continuum (again, provided $\delta > - 1$); we call this term \emph{Wilson $Q$ term}.
The Wilson $Q$ term can be chosen non-vanishing and the scheme still deliver the DE.
This observation will be useful further on, but for now let us consider that $Q = 0$. At this point one may be tempted to make the trivial choice $V=0$ to obtain $Q=0$, but this is unlikely to be possible due to the unitarity constraints  (see the Supplemental Material). The constraint we have in order to obtain $Q=0$ is $V=V^{\dag}$. We will see below what choice we finally make for $V$.

\emph{ {\bfseries Klein-Gordon-square requirement and Clifford algebra from quantum automata.}} 
Squaring the equation $\ic \mathcal{D}_0 \Psi |_j = \mathcal{H}'_{Q=0} \Psi_j$ delivers $\mathcal{D}_0 ^2 \Psi |_j = -  (\mathcal{H}'_{Q=0})^2 \Psi_j$, with $(\mathcal{H}'_{Q=0})^2 = - (A^1)^2 \mathcal{D}_1^2 +( \epsilon m)^2 (A^0)^2 + \epsilon m [ A^0 A^1 + A^1 A^0 ] (- \ic \mathcal{D}_1 )$.
Thanks to the Dirac-continuum-limit constraints, we obtain the correct Klein-Gordon equation in the continuum limit (divide by $\epsilon^2$ and let $\epsilon\rightarrow 0$ after a Taylor expansion at second order).

Now, we wish that the discrete scheme obtained by squaring Eq.\ \eqref{eq:two-step} be a valid discretization of the Klein-Gordon equation, and by valid we mean, not only that it delivers the correct continuum limit, but that it be applicable, at the discrete level, to scalar state sequences (as it is the case in the continuum setting), and for this
\emph{we need to impose $(A^0) \propto 1$ and $(A^1)^2 \propto 1$, i.e., not operator-valued}\footnote{We will see further down that requiring mere proportionality and not equality turns out to be indeed necessary, for unitarity to be preserved.}, as well as  $A^0 A^1 + A^1 A^0 \propto 1$ or equal to $0$.
\emph{We choose to impose that the $A$ operators $A^0$ and $A^1$ satisfy  $A^0 A^1 + A^1 A^0=0$}, i.e., we impose the crossed term to vanish (which may be taken as part of the definition of what we consider a valid discrete Klein-Gordon scheme). 
To sum up, what we impose is that the $A$ operators satisfy the same algebra as that of the $\alpha$ operators of the DE up to multiplicative factors, that is, we impose
\begin{subequations}
\label{eq:algebra}
\begin{align}
(A^0)^2 &\propto 1 \\
(A^1)^2 &\propto 1 \label{eq:A02}\\
A^0A^1 + A^1 A^0 &= 0 \, . \label{eq:crossed}
\end{align}
\end{subequations}

We now have to find $A$ operators which satisfy (i) this algebra, (ii) the Dirac-continuum-limit constraints, and (iii) the unitarity constraints.
Can we find such operators?
We are going to see that ``yes'', a result which is a priori not trivial at all.
The trivial choice $A^1 \propto \alpha^1$, more concretely, $B \propto \alpha^1$, is going to make us reach our purpose, so that we make this choice.
What is untrivial is the proportionality constant, as well as the choice of $M$ (which determines $A^0$).
Indeed, a standard choice for the mass term $\epsilon mA^0$  in the literature is $e^{-\ic \epsilon m \alpha^0}$, but this term together with $A^1 \propto \alpha^1$ makes Constraint \eqref{eq:crossed} unsatisfied.
The term that is going to make us reach our purpose is a suggestion from both Feynman's original scheme \cite{Schweber1986}, and Succi's quantum lattice Boltzmann schemes \cite{Succi1993, Succi2015} (while $e^{-\ic \epsilon m \alpha^0}$ is a suggestion from multiplicative constructions, while here our construction is additive): we choose $M = \mu_{\epsilon} (1 - \ic \epsilon m \alpha^0 )$, where $\mu_{\epsilon}$ is a factor that is imposed to us by one of the unitarity constraints (see the Supplemental Material), namely, $M^{\dag} M = 1$, so that $\mu_{\epsilon} \defeq \frac{1}{\sqrt{1 + \epsilon^2 m^2}}$.
This $\mu_{\epsilon}$ is also the proportionality factor evoked above: we choose $B \defeq \mu_{\epsilon} \alpha^1$.
The algebra \eqref{eq:algebra} is thus now satisfied,  since
$A^0  = \mu_{\epsilon} \alpha^0$ and $A^1  = \mu_{\epsilon} \alpha^1$, and this algebra is equivalent to the following Clifford algebra,
$ \{\Gamma^{\mu},\Gamma^{\nu}\} = 2 \tilde{\eta}^{\mu \nu}$, where $\{ \cdot, \cdot\}$ is the anticommutator, and where we have introduced the following modified Minkowski metric, $\tilde{\eta} \defeq \text{diag}(1/\mu_{\epsilon}^2,-1)$, as well as the $\Gamma$ operators,
$ \Gamma^0 \defeq (A^0)^{-1} = \alpha^0/\mu_{\epsilon}$, and $\Gamma^1 \defeq (A^0)^{-1} A^1 \defeq \ \alpha^0 \alpha^1 \defeq \gamma^1$.
One can check that the other and last unitarity constraint involving $M$ but not $V$, namely, $B^{\dag} M = M^{\dag} B$, is satisfied.

Let us recap how $M$ is built: the ``$\alpha^0$'' in $M$ is both for the unitarity constraint $B^{\dag} M = M^{\dag} B$ to be satisfied and so that $A^0A^1+A^1A^0=0$; the ``$\epsilon m$'' in front of the $\alpha^0$ is for the continuum limit to be the good one; the ``$\mathrm{i}$'' is because $A^0 \propto \mathrm{i}(M - M^{\dag})$; and finally the additional term ``$1$'' is for the unitarity constraint $M^{\dag}M=1$ to be satisfied.
Now, finally, for the unitarity constraints on $V$ to be satisfied, we can choose $V\defeq \mu_{\epsilon}$.
With these choices for $B$, $M$, and $V$, our DQW with $Q=0$ is invariant under unitary transformations of the coin state, in such a way that the algebra \eqref{eq:algebra} is preserved, in exact parallel with the continuum situation.

Note that the operators $A^0$ and $A^1$ depend on the mass, while in the continuum limit none of the $\alpha$ operators do.
Note also that $\mathcal{H}'_{Q=0} = \mu_{\epsilon} [\alpha^1 (-\mathrm{i}\mathcal{D}_1) + \epsilon m \alpha^0 ]$ (and
$(\mathcal{H}'_{Q=0})^2 = \mu_{\epsilon}^2 [- \mathcal{D}_1^2 +  (\epsilon m)^2] $), with $\mu_{\epsilon} \neq 1$ if $\epsilon m \neq 0$ (although $\mu_{\epsilon} \rightarrow 1$ as  $\epsilon m \rightarrow 0$), which is the price to pay to obtain a unitary discretization while having nevertheless discretizing the transport term naively, by a symmetric finite difference $\mathcal{D}_1$.
Finally, let us mention the following: in the Supplemental Material, we have only written the unitarity constraints ensuing from $\mathcal{U}^{\dag} \mathcal{U}=1$, but one can check that the unitarity constraints ensuing from $\mathcal{U} \mathcal{U}^{\dag}=1$ are also satisfied with our choices for $B$, $M$ and $V$.

\emph{{\bfseries Towards avoiding fermion doubling with the Wilson $Q$ term.}} 
Consider the two-step Hamiltonian of Eq.\ \eqref{eq:H_final} for a non-vanishing Wilson $Q$ term; it is convenient in order to further take a continuum limit to rather consider the Hamiltonian $h \defeq \mathcal{H}/ \epsilon$ (from the continuum point of view, this is actually the correct Hamiltonian dimensionally).
The reason we have called the Wilson $Q$ term that way is because its spatial-operator part, $-\mathcal{L}$, is that of the well-known Wilson term of LGT, namely, $(r\epsilon /2) \alpha^0 (-\mathcal{L}/\epsilon^2)$ \cite{Susskind77a}, i.e., the same as ours but with $\alpha^0$ instead of $Q$.
In LGT, this term enables to avoid the so-called fermion doubling problem, which appears when discretizing naively the DE; these facts are recalled in the Supplemental Material.
It is remarkable that such a Wilson term is naturally contained in the decomposition of a generic DQW.
What $Q$ we can take in order for the fermion doubling to be avoided, while satisfying the unitarity constraints? Is this even possible? We are going to see that the answer to the last question is ``yes'', a result that is a priori not trivial at all.

\emph{{\bfseries Klein-Gordon-square requirement with the Wilson $Q$ term.}}
Assuming $A^0A^1 + A^1 A^0 = 0$, we have that
$
h^2 = -  (A^1)^2 \frac{\mathcal{D}_1^2}{\epsilon^2} + m^2 (A^0)^2 + \frac{r^2}{4 \epsilon^2} Q^2 \mathcal{L}^2 +  m \frac{r}{2 \epsilon^2} \left[ Q {A^0}+ {A^0} Q \right] (-\mathcal{L}) + \frac{r}{2 \epsilon^2} \left[ {Q}  A^1 + A^1 {Q} \right] (-\mathcal{L}) (- \ic \mathcal{D}_1)$.
Notice that the three last terms vanish in the continuum limit, so that we recover the Klein-Gordon equation.
Now, the validity of the discrete Klein-Gordon scheme, which we define, as before, as the possibility of applying the scheme to scalar state sequences, can be obtained if we require, in addition to the previous Clifford-algebra requirement,  Eqs.\ \eqref{eq:algebra}, that as $\epsilon \rightarrow 0$, (i) $Q^2 \sim 1$, and (ii) each of the two anticommutators involving $Q$ either vanishes or is proportional to the identity; these requirements suggest to use for the choice of $V$ the same trick as that used for the choice of $M$.

Hence, we choose the ansatz $V = \nu_{\epsilon} (1 + \ic\epsilon^{\rho}  r \alpha^{\lambda})$, where (i) $\alpha^{\lambda}$ is an $\alpha$ operator satisfying the usual algebra of the $\alpha$ operators of the DE, (ii) $ \nu_{\epsilon}$ is a normalization factor to be determined, and (iii) $\rho$ is an exponent to be determined.
Since we are in one spatial dimension, $\lambda$ should only take two possible values, $\lambda = 0$ and $\lambda = 1$ (since $\alpha^{\lambda}$ has been defined as an $\alpha$ operator of the DE and we are in one spatial dimension).
That being said, and we mention this since it is going to be used further down, in general terms: let us say we are in $n$ spatial dimensions (so that $\lambda$ can take values from $0$ to $n$) with a given representation of the Clifford algebra; then $\lambda$ can take an additional value $n+1$ provided that in $n+1$ spatial dimensions one can still find a representation of the Clifford-algebra which has the same dimension as the representation found in dimension $n$\footnote{Actually, this condition may be necessary only for irreducible representations: if by going from dimension $n$ to $n+1$ the dimension of irreducible representations is augmented, one may be able to simply consider in dimension $n$ a reducible representation having the same dimension as the representation of dimension $n+1$.}. We know that this is precisely the case when going from one spatial dimension to two: e.g., if $\alpha^0$ and $\alpha^1$ are represented by two Pauli matrices, then we can choose an $\alpha^{2}$ that is represented by the last Pauli matrix.
Hence, we consider that $\lambda = 0$, $1$, or $2$.

The unitarity constraint $B^{\dag}V = V^{\dag} B$ requires $\lambda \neq 1$, which we assume, so that the anticommutator $Q A^1 + A^1 Q$ vanishes.
We do not change our choice of $M$, neither its normalization factor (the latter enables to satisfy $M^{\dag} M=1$). 
Now, both $\lambda=0$ and $\lambda=2$ are compatible with the unitarity constraint $2V^{\dag}V = V^{\dag} M + M^{\dag} V$, which determines the following associated normalization factors: $ \nu_{\epsilon}^{\lambda=0} \defeq \mu_{\epsilon} \frac{1 - \epsilon^{1+\rho} m r}{1 + (\epsilon^{\rho}r)^2}$, and $\nu_{\epsilon}^{\lambda=2} \defeq \frac{\mu_{\epsilon} }{1 + (\epsilon^{\rho}r)^2}$.
Finally, for the unitarity constraint $V^{\dag} V = B^{\dag}B$ to be satisfied, we need to change the normalization factor for $B$: we choose $B \defeq \eta_{\epsilon}^{\lambda} \alpha^1$, and $\eta_{\epsilon}^{\lambda}$ is determined by the constraint just mentioned, yielding
$\eta_{\epsilon}^{\lambda} \defeq \nu_{\epsilon}^{\lambda} \sqrt{1 + (\epsilon^{\rho}r)^2}$.
All unitarity constraints are now satisfied, including those coming from $\mathcal{U}\mathcal{U}^{\dag}=1$.
Now, since $\mu_{\epsilon}$ goes to $1$ in the continuum limit, we see that for both $\nu^{\lambda} _{\epsilon}$ and $\eta^{\lambda} _{\epsilon}$ to go to $1$ in the continuum limit, we must choose $\rho > 0$:
this finally yields $Q = \nu_{\epsilon} \epsilon^{\rho} \alpha^{\lambda}$, which in the continuum limit goes as $\epsilon^{\rho} \alpha^{\lambda}$ -- that is, in the case $\lambda=0$, as $\epsilon^{\rho} \alpha^{0}$ and not as $\alpha^0$ as in standard LGT.
We finally have
\begin{equation}
\label{eq:final_hh}
h^{\lambda} = \eta_{\epsilon}^{\lambda} \alpha^1 \Big( \frac{- \ic \mathcal{D}_1}{\epsilon} \Big) +  \mu_{\epsilon} m \alpha^0 +\nu^{\lambda}_{\epsilon} \epsilon^{\rho} \frac{r}{2 \epsilon} ( - \mathcal{L} ) \alpha^{\lambda}      \, .
\end{equation}
We will keep working with both models $\lambda=0$ and $\lambda=2$ and see if one performs better than the other at avoiding fermion doubling (avoidance which for now is not a given, we are precisely going to explain further down at which condition fermion doubling is avoided).

\emph{{\bfseries Avoiding fermion doubling.}}
To find solutions of our two-step scheme, Eq.\ \eqref{eq:two-step}, we consider a superposition-of-plane-waves ansatz (since the solution of the DE, that we seek to simulate, has this form):
$\Psi(t,x) = \frac{1}{2\pi N} \sum_{i=+,-} \int_{-\infty}^{+\infty} dk \, \tilde{\Psi}_i(0,k)  e^{-\ic  ( \omega_i(k) t - k x)}$, where $(t,x) = (t_j,x_p)$.
Inserting the plane-wave ansatz into Eq.\ \eqref{eq:two-step} with $h = \mathcal{H}/\epsilon$ given by Eq.\ \eqref{eq:final_hh}, leads, after squaring, to a dispersion relation
\begin{equation}
\frac{\sin^2(\omega_{i}(k) \epsilon)}{\epsilon^2} = F^{\text{DQW},\lambda}(k) \, ,
\end{equation}
where
{\footnotesize
\begin{subequations}
\label{eqs:F}
\begin{align}
F^{\text{DQW},\lambda=0}(k) &\defeq ( \eta_{\epsilon}^0)^2 \frac{\sin^2(k \epsilon)}{\epsilon^2} + \Big[ \mu_{\epsilon} m + \nu_{\epsilon}^0 \epsilon^{\rho}  \frac{r}{\epsilon} \big(1 - \cos (k \epsilon) \big) \Big]^2 \label{eq:F1}\\
F^{\text{DQW},\lambda=2}(k) &\defeq  (\eta_{\epsilon}^2)^2 \frac{\sin^2(k \epsilon)}{\epsilon^2} +  (\mu_{\epsilon} m)^2 + \Big[ \nu_{\epsilon}^2 \epsilon^{\rho}  \frac{r}{\epsilon} \big(1 - \cos (k \epsilon) \big) \Big]^2 \, . \label{eq:F2}
\end{align}
\end{subequations}}

Now, discrete-time LGT is usually formulated in a Lagrangian way \cite{book_Rothe, Hernandez2011}, whereas we formulated our discrete-time scheme in a Hamiltonian one.
In Lagrangian LGT, a term is naturally added to the action to remove the temporal doublers along with the term added to remove the spatial doublers.
Here, this is not the case, and we only treat spatial doublers.
Considering a low-frequency limit, $\omega_{i}(k) \ll 1$, of the dispersion relation, finally leads to the solutions
\begin{equation}
\omega^{\text{DQW},\lambda}_{\pm}(k) = \pm \sqrt{F^{\text{DQW}, \lambda}(k) } \, .
\end{equation}
Expressions \eqref{eqs:F} are to be compared with the dispersion relation of standard LGT,
\begin{equation}
\label{eq:FLGT}
F^{\text{LGT}}(k) \defeq \frac{\sin^2(k \epsilon)}{\epsilon^2} + \Big[  m +   \frac{r}{\epsilon} \big(1 - \cos (k \epsilon) \big) \Big]^2 \, .
\end{equation}

The difference between $F^{\text{DQW},\lambda=0}(k)$ and $F^{\text{DQW},\lambda=2}(k)$ is, apart from the normalization factors $\nu^{\lambda}_{\epsilon}$ and  $\eta^{\lambda}_{\epsilon}$, the \emph{crossed term}  of the square of the sum in $F^{\text{DQW},\lambda=0}(k)$.
In standard LGT, this crossed term is also present.
Since the upcoming discussion is the same whether there is a crossed term or not, let us forget about the latter for now, we will come back to it later.
Here comes the discussion.
In standard LGT, what makes fermion doubling be avoided is that the function $1-\cos(k\epsilon)$ raises the value of the dispersion relation at the edges of the Brillouin zone, i.e., at $\pm\pi/\epsilon$.
More precisely, if the amplitude by which this dispersion relation is raised, which we call \emph{raising amplitude}, does not go to zero with $\epsilon$, then for sure fermion doubling is avoided, and this is indeed the case in standard LGT since the associated raising amplitude is $(r/\epsilon)^2$, see Eq.\ \eqref{eq:FLGT}.
In our model, the raising amplitude is $(\nu^{\lambda}_{\epsilon} r \epsilon^{\rho - 1})^2$, see Eqs.\ \eqref{eqs:F}.
Since $\nu^{\lambda}_{\epsilon}$ goes to $1$ in the continuum limit, we are fine as long as
\begin{equation}
\rho < 1 \, ,
\end{equation}
i.e., this condition ensures that fermion doubling is avoided, both for $\lambda = 0$ and $\lambda = 2$.

\emph{{\bfseries Initial slopes.}}
The limit to the continuum is obtained by letting $\epsilon$ go to zero. Let us consider the gapless frequencies
\begin{equation}
\label{eq:gapplessfreq}
f^{\text{M}}(k) \defeq \sqrt{ (\omega^{\text{M}}_{\pm}(k))^2 - M^{\text{M}} } = \sqrt{ F^{\text{M}}(k) - M^{\text{M}} } \, ,
\end{equation}
where ``$\text{M}$'' is the considered model, and takes for now the ``values'' $\text{M}= (\text{DQW}, \lambda=0)$ or $\text{M}= (\text{DQW}, \lambda=2)$, and where $M^{\text{M}}$ is the central $(k=0)$ gap of the model. Let us consider the Taylor expansion of the dispersion relations $(f^{\text{M}}(k))^2$ at next-to-lowest order in $\epsilon$. A priori, the two Taylor-expanded dispersion relations (one for $\lambda=0$ and one for $\lambda = 2$) are different: more precisely, there are a priori more terms in the case $\lambda=0$, namely, the crossed terms. Now, it turns out that these additional terms actually cancel each other, at least at next-to-lowest order in $\epsilon$, so that both the model $\lambda=0$ and $\lambda=2$ have, at next-to-lowest order in $\epsilon$, the same Taylor expansion in $\epsilon$, namely,
\begin{equation}
\label{eq:slope}
(f^{\text{DQW},\lambda=0}(k))^2 \simeq (f^{\text{DQW},\lambda=2}(k))^2 \simeq (1-\frac{1}{2} r^2 \epsilon^{2\rho})k^2 \, .
\end{equation}
Actually, even if these above-mentioned additional terms had not cancelled each other, they would be of higher order because $\rho < 1$, so both models $\lambda = 0$ and $\lambda=2$ would still have the same small-$\epsilon$ expansion, we leave the verification of this observation to the reader.
Moreover, the criterion $\rho < 1$ for the avoidance of fermion doubling in our model, is the same whether there is a crossed term or not.
Hence, we consider both models $\lambda = 0$ and $\lambda = 2$ equivalently good for our task of avoiding fermion doubling and converging as fast as possible to the continuum limit.
We choose $\lambda = 0$, in order for comparisons to LGT to be simpler, and from now on we call this model $\text{M} = \text{DQW}$.
At this point the reader may wonder why we have kept the discussion about $\lambda=0$ and $\lambda=2$ if in the end we consider both models equivalently good and choose e.g. $\lambda=0$: apart from a mere informative reason, this discussion is going to be useful  just below in the next paragraph, to understand the role of the crossed term in the LGT model at small $\epsilon$, and the subsequent interest of getting rid of this crossed term by choosing $\lambda = 2$ in this LGT model.

We call the factor in front of $k^2$ in Eq.\ \eqref{eq:slope} the (squared) initial slope of the model.
In standard LGT, the lowest-order modification of the initial slope comes, this time, from the crossed term (this is detailed in the Supplemental Material): at next-to-lowest order in $\epsilon$ we have that
\begin{equation}
\label{eq:slopeLGT}
(f^{\text{LGT}}(k))^2 \simeq (1 + \epsilon m r)k^2 \, .
\end{equation}
Hence, for the initial slope of our model to converge faster to the continuum limit than that of the LGT model, we need to choose
\begin{equation}
\rho > 0.5 \, .
\end{equation}
In the end, we have, in addition to in-built unitarity, a model that performs better at reaching the continuum limit than that of LGT if we choose $\rho \in \ ]0.5,1[$.
That being said, if we choose in the LGT model $\lambda = 2$, then the initial slope is exactly that of the continuum, that is, $1$ (we leave the verification of this to the reader), and in that case the LGT model performs better than ours at reaching the continuum limit, whatever value we choose for $\rho$. 

\begin{figure}[h!]
\includegraphics[width=8.9cm]{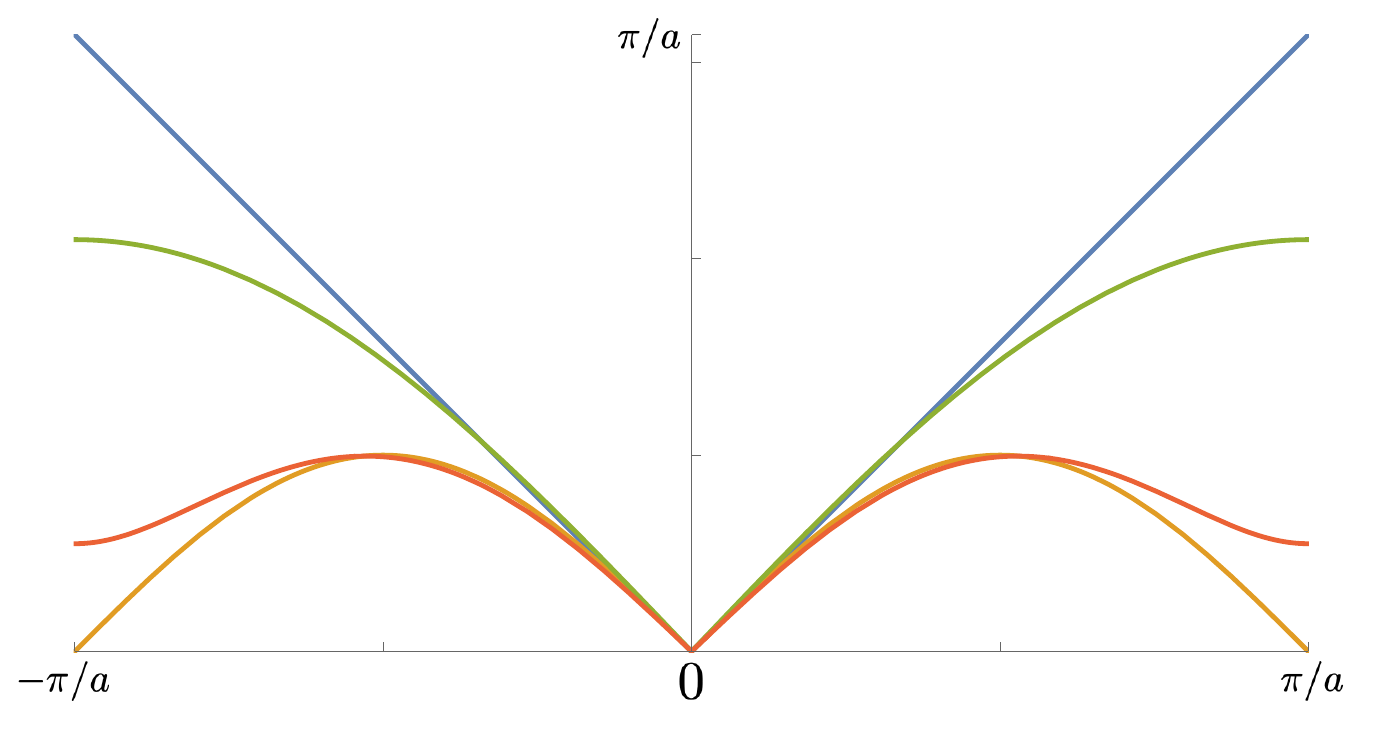}
\vspace{-0.9cm}
\caption{Gapless frequency $f^{\text{M}}(k)$ of Model $\text{``M''}$, for $\text{``M''} = \text{``Dirac''}$ (blue), $\text{``naive''}$ (gold), $\text{``LGT''}$ (green), and $\text{``DQW''}$ (red), where  $f^{\text{Dirac}}(k) = |k|$, $f^{\text{naive}}(k) = \tfrac{|\sin(k\epsilon)|}{\epsilon}$,  $f^{\text{LGT}}(k) = \sqrt{\tfrac{\sin^2(k\epsilon)}{\epsilon^2} + \big[m + \tfrac{r}{\epsilon} (1 - \cos(k\epsilon)\big]^2 -m^2}$, and $f^{\text{DQW}}(k) = \sqrt{(\eta^{0}_{\epsilon})^2\tfrac{\sin^2(k\epsilon)}{\epsilon^2} + \big[\mu_{\epsilon}m +  \nu^{0}_{\epsilon} \epsilon^{\rho}  \tfrac{r}{\epsilon} (1 - \cos(k\epsilon)\big]^2 - (\mu_{\epsilon} m)^2}$, for $\epsilon = 0.1$, $m=1$, $r=1$ (Wilson's choice), and $\rho = 0.6$. \label{fig:Figure}}
\end{figure}

\emph{{\bfseries Figure.}}
In  Fig.\ \ref{fig:Figure}, we plot the gapless frequencies defined above in Eq.\ \eqref{eq:gapplessfreq}, where this time the model $\text{M}$ takes the ``values'' $\text{M}= \text{Dirac}, \text{naive}, \text{LGT}, \text{DQW}$.
We see that the naive discretization of the DE leads to two extra poles in the gapless frequency (gold plot), which causes the fermion doubling problem (see the Supplemental Material), while the LGT (green plot) and DQW (red plot) models avoid it by creating gaps at the edges.
In-built unitarity \emph{for any $r \in \mathbb{R}$} is an important advantage of our model with respect to LGT.

Take a closer look at the initial slopes (i.e., around $k=0$).
One can appreciate, though weakly, the fact that the initial slope of the DQW (resp. LGT) model is a bit smaller, see Eq.\ \eqref{eq:slope} (resp. bigger, see Eq.\ \eqref{eq:slopeLGT}) than that of Dirac fermions (although the two former slopes obviously converge to the latter in the continuum limit i.e. for $\epsilon$ going to $0$).
We could have made this fact more visible on the figure, but then the plots do not give the impression (which is anyways of course a truth, as shown above) that in the continuum limit one does get the correct initial slope, while in Fig.\ \ref{fig:Figure} this impression is given.

Let us make a final remark. In the figure, we have chosen $r=1$ because it is a standard choice. That being said, let us mention that for $r=1$ the discrete-time LGT model of Wilson fermions (which is a Lagrangian model) can actually be proven to be unitary. This proof does however not hold for $r \neq 1$ \cite{Hernandez2011}, while our model has in-built unitarity for any $r$. Pay attention that if we choose, for $r=1$, $\lambda=2$ in the discrete-time LGT model, unitarity would again have to be proven, and it may actually not hold.

%

\emph{{\bfseries Conclusion.}}
We showed that a Clifford algebra emerges out of the internal-state operators defining a quantum-automaton discretization of the Dirac equation.  This discretization, which is a DQW, is unitary by construction, while discrete-time versions of lattice gauge theory are usually Lagrangian so that unitarity is not in-built. Our DQW is invariant under representation changes of the above-mentioned Clifford algebra, which parallels exactly the continuum situation. Moreover, we show that DQWs naturally contain a Wilson term that makes spatial fermion doubling be avoided, and this without breaking unitarity and for any choice $r \in \mathbb{R}$ of Wilson's parameter.

\emph{{\bfseries Acknowledgements.}}
The author thanks C. Cedzich for making him notice that the two-step Hamiltonian is (ultra)local, while the effective Hamiltonian, as a logarithm of the one-step evolution operator, is not (infinite series of growing powers of the one-step evolution operator).
This is an interesting feature which is in the prolongation of the ultralocality of the one-step evolution operator that is seeked in the development of quantum cellular automata and their one-particle versions, DQWs.
The author also thanks A. Grinbaum for his stylistic and grammatical energetic advice and corrections.


\let\oldaddcontentsline\addcontentsline
\renewcommand{\addcontentsline}[3]{}
\let\addcontentsline\oldaddcontentsline

\clearpage
\newpage

\begin{widetext}

\begin{center}
{\bfseries \large Supplemental Material}
\end{center}

\noindent
{\small In the first section, we translate, on the transport coin operators $B$, $V$ and $M$, the unitarity constraints imposed on the jump coin operators $W_i$, $i=-1,0,1$, by the unitarity of the one-step evolution operator $\mathcal{U}$. In the second section, we explain the fermion-doubling problem of lattice gauge theory, and present Wilson's method to solve it, with so-called Wilson fermions.}

\tableofcontents

\section{Unitarity constraints}

We impose the unitarity of $\mathcal{U}$, that is, $\mathcal{U}^{\dag} \mathcal{U} = 1$.
A couple of computation lines lead to the following conditions on the jump coin operators $W_i$, $i=-1,0,1$, that we call unitarity constraints on the jump coin operators:
\begin{subequations}
\begin{align}
W_{-1}^{\dag}W_{-1} + W_{1}^{\dag}W_{1}  + W_{0}^{\dag}W_{0} &= 1 \label{eq:constraintA} \\
W_{-1}^{\dag}W_{0}  + W_{0}^{\dag}W_{1} &= 0 \label{eq:constraintB} \\
W_{-1}^{\dag}W_{1} &= 0 \, . \label{eq:constraintC} 
\end{align}
\end{subequations}

Let us  translate these constrains on the transport coin operators $B$, $V$ and $M$.
We start by Constraint \eqref{eq:constraintC}, which is the simplest to translate in the sense that it does not involve $M$ while both Constraints  \eqref{eq:constraintA} and  \eqref{eq:constraintB} do.
Constraint \eqref{eq:constraintC} yields
\begin{subequations}
\begin{align}
0 &= (V^{\dag} - B^{\dag}) (V+B) \\
&= V^{\dag}V -  B^{\dag}B - B^{\dag}V + V^{\dag}B \, . \label{eq:constraintCprime}
\end{align}
\end{subequations}
Computing the sum and the difference, $\text{\eqref{eq:constraintCprime}}^{\dag} + \text{\eqref{eq:constraintCprime}}$ and $\text{\eqref{eq:constraintCprime}}^{\dag} - \text{\eqref{eq:constraintCprime}}$ yields, respectively,
\begin{subequations}
\begin{empheq}[box=\widefbox]{align}
V^{\dag}V &= B^{\dag} B  \label{eq:constraint1} \\
B^{\dag}V &=  V^{\dag} B \, . \label{eq:constraint2}
\end{empheq}
\end{subequations}
Conversely, it is trivial to check that these two constraints imply Constraint \eqref{eq:constraintC}.

We now proceed with translating Constraint \eqref{eq:constraintB}, because we will use one of the two resulting constraints to translate Constraint \eqref{eq:constraintA}.
Constraint \eqref{eq:constraintB} yields
\begin{subequations}
\begin{align}
0 &= (V^{\dag} - B^{\dag}) (M-V)  +(M^{\dag}-V^{\dag}) (V+B) \\
&= V^{\dag}M - B^{\dag} M - V^{\dag} V + B^{\dag} V + M^{\dag} V - V^{\dag}V + M^{\dag}B - V^{\dag} B \, . \label{eq:constraintBprime}
\end{align}
\end{subequations}
Computing the sum and the difference $\text{\eqref{eq:constraintBprime}}^{\dag} + \text{\eqref{eq:constraintBprime}}$ and $\text{\eqref{eq:constraintBprime}}^{\dag} - \text{\eqref{eq:constraintBprime}}$,  and using Eq.\ \eqref{eq:constraint2} in the difference, yields, respectively,
\begin{subequations}
\begin{empheq}[box=\widefbox]{align}
2V^{\dag}V &= V^{\dag} M + M^{\dag} V  \label{eq:constraint3} \\
B^{\dag} M &=  M^{\dag} B \, . \label{eq:constraint4}
\end{empheq}
\end{subequations}
Conversely, it is trivial to check that these two constraints together with Constraint \eqref{eq:constraint2}  imply Constraint \eqref{eq:constraintB}.

We finally proceed with translating Constraint \eqref{eq:constraintA}, which can be written
\begin{subequations}
\begin{align}
4 &=  (V^{\dag} - B^{\dag}) (V-B) + (V^{\dag} + B^{\dag}) (V + B)  + 4 W_0^{\dag} W_0 \\
&= V^{\dag}V + B^{\dag} B - B^{\dag} V - V^{\dag} B + V^{\dag} V + B^{\dag} B + B^{\dag} V + V^{\dag} B + 4 W_0^{\dag} W_0   \, . \label{eq:constraintAprime}
\end{align}
\end{subequations}
Inserting both Constraint \eqref{eq:constraint1} and Constraint \eqref{eq:constraint2} in the preceding one, Constraint \eqref{eq:constraintAprime}, delivers
\begin{subequations}
\begin{align}
1 &= V^{\dag} V + W_0^{\dag} W_0  \\
&= V^{\dag} V + (M^{\dag}-V^{\dag}) (M-V) \\
&= 2 V^{\dag} V + M^{\dag} M - M^{\dag} V - V^{\dag} M \, . \label{eq:constraintAprimeprime}
\end{align}
\end{subequations}
Finally, inserting Constraint \eqref{eq:constraint3} into the preceding one, Constraint \eqref{eq:constraintAprimeprime}, results in
\begin{empheq}[box=\widefbox]{equation}
M^{\dag} M = 1 \, .
\end{empheq}

\section{Fermion doubling and Wilson fermions in continuous and discrete time}

\subsection{The Schrödinger equation for translationally invariant systems}
\label{subsec:Schro}

\subsubsection{The generic Schrödinger equation and its spectral solution}

The generic Schrödinger equation is a partial differential equation (PDE) of the form
\begin{equation}
\label{eq:Schrodinger}
\ic \partial_0 \Psi |_{t} = h \Psi(t) \, ,
\end{equation}
where $h$ is a Hermitean linear operator acting on the function $\Psi(t) : x \mapsto \Psi(t,x)$, where here we choose $x \in \mathbb{R}$.
Since this equation is linear, we solve it spectrally, i.e., by finding the eigen-elements $(\omega_{\sigma},\Phi_{\sigma})_{\sigma \in \Sigma}$ of $h$, where $\Sigma$ is a certain indexing space; By definition, these eigen-elements satisfy
\begin{equation}
\label{eq:eigenelements}
h \Phi_{\sigma} = \omega_{\sigma} \Phi_{\sigma} \, ,
\end{equation}
where the eigen-values $\omega_{\sigma}$ are real since $h$ is Hermitean. 

The method is the following.
Assume we have determined the eigen-elements of $h$.
Since $\Psi(t)$  belongs to a Hilbert space, we can decompose it on the eigen-basis $(\Phi_\sigma)_{\sigma \in \Sigma}$ at an arbitrary time $t$:
\begin{equation}
\Psi(t) = \sum_{\sigma \in \Sigma} C_{\sigma}(t) \Phi_{\sigma} \, ,
\end{equation}
where the $C_{\sigma}(t)$s are the coefficients of $\Psi(t)$ on the eigen-basis.
Now, using Eq.\ \eqref{eq:eigenelements}, the generic Schrödinger equation on $\Psi$, Eq.\ \eqref{eq:Schrodinger}, which is a PDE, can be translated into a family of ordinary differential equations (ODEs) -- indexed by $\sigma$ -- for the coefficients $C_{\sigma}$, that is,
\begin{equation}
\ic \partial_0 C_{\sigma}|_{t} = \omega_{\sigma}  C_{\sigma}(t) \, ,
\end{equation}
whose solution is well-known:
\begin{equation}
C_{\sigma}(t) = C_{\sigma}(0) e^{-\ic \omega_{\sigma} t} \, .
\end{equation}
Hence, the solution seeked is
\begin{empheq}[box=\widefbox]{equation}
\label{eq:solution}
\Psi(t) = \sum_{\sigma \in \Sigma} C_{\sigma}(0) e^{-\ic \omega_{\sigma} t}  \Phi_{\sigma} \, .
\end{empheq}
Because they intervene in the periodic functions $t \mapsto e^{-\ic \omega_{\sigma} t}$, the $\omega_{\sigma}$ are called \emph{frequencies}; More precisely, they are the \emph{eigen-frequencies of $h$}. To be more definite, one should actually use the denomination ``angular frequency'' rather than ``frequency''.

\subsubsection{Fourier analysis}

We will use the more definite notation $\Psi(t,\cdot)$ for $\Psi(t)$ when needed.
Let us take the Fourier transform of $\Psi(t,\cdot)$ at a given time $t$:
\begin{equation}
\tilde{\Psi}(t,k) \defeq \frac{1}{\sqrt{2 \pi}} \int_{\mathbb{R}} dx \, \Psi(t,x) e^{-\ic k x}  \, .
\end{equation}
Inverting this equation, we obtain the decomposition of $\Psi(t)$ into its Fourier components:
\begin{equation}
\label{eq:Fourier_components}
\Psi(t,x) =  \frac{1}{\sqrt{2 \pi}} \int_{\mathbb{R}} dk \, \tilde{\Psi}(t,k) e^{\ic k x} \, .
\end{equation}
To be precise, the function $x \mapsto \tilde{\Psi}(t,k) e^{\ic k x}$ is the \emph{Fourier component} of $\Psi(t,\cdot)$ associated to the value $k$ of the Fourier variable, and $\tilde{\Psi}(t,k)$ is the \emph{Fourier coefficient}, or \emph{Fourier amplitude} of $\Psi(t,\cdot)$ associated to the value $k$.
Because $x$ is a spatial position, the Fourier variable $k$ is a \emph{spatial frequency}; Again, as in the case of $\omega_{\sigma}$ above, to be more definite one should use the denomination ``angular pulsation'' rather than ``frequency''.

\subsubsection{Case of translationally invariant systems}

If $h$ does not depend on the point $x$, i.e., if $h$ is translationally invariant, then one can check by considering Eq.\ \eqref{eq:Schrodinger} for Expression \eqref{eq:Fourier_components} that each Fourier coefficient satisfies the equation
\begin{equation}
\label{eq:Schrodinger_in_Fourier}
\ic \partial_0 \tilde{\Psi}(\cdot,k)|_{t} = \tilde{h}(k) \tilde{\Psi}(t,k) \, ,
\end{equation}
where $\tilde{h}(k)$ is the expression obtained when replacing, in $h$, the operator $-\ic \partial_1$ by the real number $k$.

Hence, if $h$ does not depend on the point $x$, each Fourier coefficient evolves independently of the others, while this is not the case if $h$ does depend on $x$. 
Moreover, the ``Schrödinger equation in Fourier space'', Eq.\ \eqref{eq:Schrodinger_in_Fourier}, is simpler to solve than the original  Schrödinger equation, \eqref{eq:Schrodinger}, because the operator $-\ic \partial_1$ has been replaced by a real number $k$, so that Eq.\ \eqref{eq:Schrodinger_in_Fourier} is not a PDE as the original Schrödinger equation, but a \emph{family} of ODEs indexed by $k$.
Let us now make the link with Sec.\ \ref{subsec:Schro}: One can actually mathematically show (via, e.g., ``mere'' constructive proofs) for a large class of operators $h$, that there exist an indexing space $\Sigma$ such that $k$ is one of the indices, i.e., $k \in \sigma$ \footnote{We take the liberty to consider $\sigma$ as convenient; Here, it is a set of indices, i.e., an unordered family; In practice, we will often order its elements, i.e., we will consider a list rather than a set.}.
In the language of quantum mechanics, and index $i \in \sigma$ is refered to as a good quantum number; It is an eigenvalue of an operator whose diagonalization serves as a partial diagonalization of $h$, i.e., $h$ is codiagonalizable with that operator.

We often speak of \emph{Fourier modes} for the Fourier components; The term ``mode'' refers, in its most general acception, to one of the terms of a particularly relevant decomposition of an object, take, e.g., the Fourier decomposition of a function.
In the case of an $x$-independent $h$, the Fourier modes are actually \emph{proper modes}, because by considering the Fourier version $\tilde{h}(k)$ of $h$ we have (at least partially) ``diagonalized''  $h$ (which is summed up by writing $k \in \sigma$).

\subsection{The solution of the Schrödinger equation for translationally invariant systems: A superposition of plane waves}
\label{subsec:planewaves}

\subsubsection{No internal structure for $\Psi(t,x)$}

\paragraph{Final solution.} If $\Psi(t,x)$ has no internal structure, i.e., if $\Psi(t,x) \in \mathbb{C}$, then we simply have that $k = \sigma$, and the eigen-values are $\omega_{\sigma} = \omega(k) \defeq \tilde{h}(k)  \in \mathbb{R}$; The sum over $\sigma$ in Eq.\ \eqref{eq:solution} is an integral over $k$, and the eigen-basis is the following family of functions of $x$, $(\Phi(\cdot,k))_{k\in \mathbb{R}}$, with
\begin{equation}
\label{eq:plane-wave}
\Phi(x,k) \defeq \frac{e^{\ic k x}}{\sqrt{2 \pi}} \, ,
\end{equation}
so that the solution given in Eq.\ \eqref{eq:solution} here reads
\begin{equation}
\label{eq:almost_final_solution}
\Psi(t,x) = \frac{1}{\sqrt{2\pi}} \int_{\mathbb{R}}  dk \,  C(0,k) \, e^{-\ic (\omega(k) t - k x)}   \, .
\end{equation}
By taking $t=0$ in this equation we realize by identification (they are unique) that the $C(0,k)$ are the Fourier coefficients of $x \mapsto \Psi(0,x)$, i.e.,
\begin{equation}
C(0,k) \equiv \tilde{\Psi}(0,k) \, .
\end{equation}

The value $\omega(k)$ is the eigen-frequency associated to the spatial frequency $k$.
Now, an important remark is that $\Psi(t,\cdot) $ is actually a supersposition of plane waves, with weights the Fourier coefficients of the initial condition.
Because of this $\text{(plane-)wave}$ structure, the spatial frequency $k$ is called \emph{wavevector}\footnote{The question that one may ask oneself is whether there exist a ``physical'' $h$ that is both linear and Hermitean but that does not lead to an ondulatory phenomenology.}.
The fact that the frequency of the wave $\omega(k)$ depends on the wavevector $k$ is called \emph{dispersion}, and the expression $\omega(k)$ is called the \emph{dispersion relation}.
Notice that the fact that we have a dispersion phenomenon while we are in vacuum is specific to quantum mechanics, more precisely, to the quantum mechanics of massive bodies.

In quantum mechanics, the use of the word ``wavevector'' is actually extended to non translationally invariant systems, because this wavevector is, due to the wave-particle duality, in one-to-one correspondence with the momentum of the free particle associated to the wave in question, solution of a free (i.e., with $x$-independent $h$) Schrödinger equation: More precisely, $k$ is the De Broglie wavevector of a particle of momentum $p = k$ (with $\hbar = 1$).
There is also an analog relation for the eigen-frequencies, refered to as Einstein's relation: The frequency $\omega_{\sigma}$ is the frequency associated to an energy $E_{\sigma} = \omega_{\sigma}$ (with $\hbar = 1$) for the particle in question.

\paragraph{Final solution rederived by focusing on the linear algebraic structure.} It is the opportunity to rederive the above solution, Eq.\ \eqref{eq:almost_final_solution}, by focusing on the linear algebraic structure of the computations.
This will make gentler the above replacement of $\sigma$ by $k$.
We first rewrite Eq.\ \eqref{eq:solution} with a braket notation, which emphasizes the linear algebraic structure,
\begin{equation}
\label{eq:solution2}
\ket{\Psi(t)} = \sum_{\sigma \in \Sigma} \langle \underbrace{\Phi_{\sigma} | \Psi(0) \rangle}_{\equiv \, C_{\sigma}(0)} e^{-\ic \omega_{\sigma} t}  \ket{\Phi_{\sigma}} \, .
\end{equation}
By taking $t=0$ in this equation, we see that we have simply applied the closure relation $\sum_{\sigma \in \Sigma} \ket{\Phi_{\sigma}} \! \! \bra{\Phi_{\sigma}}$ to $\ket{\Psi(0)}$, and then evolved the eigen-states up to time $t$.
We then consider the present case $\sigma = k$, so that $\Phi_{\sigma} = \Phi(\cdot, k)$, which means that the eigen-basis is the basis made up of the $\ket{k} \defeq \ket{\Phi (\cdot,k)} $ with $k \in \mathbb{R}$,
\begin{equation}
\label{eq:solution3}
\ket{\Psi(t)} = \int_{\mathbb{R}} dk \, \langle \underbrace{k | \Psi(0) \rangle}_{\equiv \,  \tilde{\Psi}(0,k)} e^{-\ic \omega_{\sigma} t}  \ket{k} \, .
\end{equation}
We then apply the bra $\bra x$ to have the value at point $x$,
\begin{equation}
\label{eq:solution4}
\langle x | \Psi(t) \rangle = \int_{\mathbb{R}}  dk \, \tilde{\Psi}(0,k)  e^{-\ic \omega_{\sigma} t}  \langle x | \Phi(\cdot,k) \rangle \, ,
\end{equation}
that is, using Eq.\ \eqref{eq:plane-wave}, exactly Eq.\ \eqref{eq:almost_final_solution},
\begin{empheq}[box=\widefbox]{equation}
\Psi(t,x) = \frac{1}{\sqrt{2\pi}} \int_{\mathbb{R}}  dk \,  \tilde{\Psi}(0,k)  e^{-\ic  ( \omega(k)t - k x)} \, .
\end{empheq}

\subsubsection{Internal structure for $\Psi(t,x)$}

Now, if $\Psi(t,x)$ has an internal structure, then $k \varsubsetneq \sigma$ \footnote{We have identified $k$ with the singlet $\{ k \}$.}, and $\tilde{h}(k)$ can be seen as a matrix indexed by $k$, that one has to diagonalize to finally find the eigen-values of $h$. 
The eigen-values of $\tilde{h}(k)$ can be denoted  $\omega_{\sigma} = \omega_{i}(k)$, where $i = 1, ..., d$, with $d$ the dimension of the matrix $\tilde{h}(k)$ (some eigen-values may be equal, e.g., $\omega_{i_a}(k) = \omega_{i_b}(k)$).
The final solution is then
\begin{empheq}[box=\widefbox]{equation}
\label{eq:The_solution}
\Psi(t,x) = \sum_{i = 1}^d \frac{1}{\sqrt{2\pi} N} \int_{\mathbb{R}}  dk \,  \tilde{\Psi}_i(0,k)  e^{-\ic  ( \omega_i(k) t - k x)} \, ,
\end{empheq}
where $N$ is a normalization factor, needed if we want the eignevectors $ \tilde{\Psi}_i(0,\cdot) $ of $\tilde{h}(k)$ to be normalized.
We see that the solution is still a superposition of plane waves, and there are $d$ dispersion relations $\omega_i(k)$, $i=1, ..., d$.

\subsection{Dirac fermions (continuous spacetime)}
\label{subsec:Dirac}

The Dirac Hamiltonian in $1+1$ dimensions is
\begin{equation}
\hat{h}^{\text{Dirac}} \defeq \alpha^1 \hat{k} + m \alpha^0 \, ,
\end{equation}
where $\hat{k}$ is the momentum operator, i.e., the abstract version of the operator $ - \ic \partial_1$, and the alpha matrices satisfy $(\alpha^0)^2= (\alpha^1)^2 = 1$ and $\alpha^0 \alpha^1 + \alpha^1 \alpha^0 = 0$.
We see that $\hat{h}^{\text{Dirac}}$ is Hermitean, and so it is a valid Hamiltonian for the generic Schrödinger equation considered in Sec.\ \ref{subsec:Schro} and \ref{subsec:planewaves} above.
Moreover it is translationally invariant.

We consider the ``Hamiltonian in momentum space'',
\begin{equation}
\tilde{h}^{\text{Dirac}}(k)  \defeq \alpha^1 k + m \alpha^0
\end{equation}
The eigen-value equation with unknowns the eigen-elements $(\omega_i(k),V_i(k))_{i=1, ..., d; k \in \mathbb{R}}$ of $ \tilde{h}^{\text{Dirac}}(k)$ is, in matrix notation,
\begin{equation}
\label{eq:eigen-equation}
\tilde{h}^{\text{Dirac}}(k) V_i(k) = \omega_i(k) V_i(k) \, .
\end{equation}

Now, to find the eigen-values of $\tilde{h}^{\text{Dirac}}(k)$, there is actually a ``trick'', related to the fact that the square of the Dirac equation is, by (historical) construction of the Dirac equation, the Klein-Gordon equation, and hence applicable to scalar state functions: The square $\big(\tilde{h}^{\text{Dirac}}(k)\big)^2$ is proportional to the identity matrix. 
By squaring Eq.\ \eqref{eq:eigen-equation}, we arrive to
\begin{equation}
\omega(k)^2 = k^2 + m^2 \, ,
\end{equation}
so that the eigen-values are
\begin{empheq}[box=\widefbox]{equation}
\label{eq:Dirac_disp}
\omega_{\pm}^{\text{Dirac}}(k) \defeq \pm \sqrt{k^2 + m^2} \, .
\end{empheq}

\subsection{The doubling problem when discretizing space but keeping time continuous: Spatial doublers}
\label{subsec:continuous_time}

To discretize space, we can simply perform the naive replacement of the partial derivative $\partial_1$ by a finite difference on a $1\text{D}$ lattice that we introduce, with sites labeled by $p \in \mathbb{Z}$ and lattice spacing $a$.
The finite difference has to be symmetric if we want the resulting Hamiltonian to be Hermitean.
Since the translation operator in the direction of growing $p$s is, in abstract space, $\hat{T} = e^{-\ic \hat{k} a}$ \footnote{Pay attention to the fact that, although we use the notation $\hat{k}$, on the lattice $\hat{k}$ and the position operator $\hat{x}$ do not satisfy the canonical commutation relations.}, the substitution of $\partial_1 = \ic \mathcal{K}$, where $\mathcal{K} \defeq - \ic \partial_x$, by the announced symmetric finite difference, 
\begin{equation}
\mathcal{D}_1 \defeq \frac{1}{2} (\T^{-1} - \T) \, ,
\end{equation}
corresponds to the following substitution in the ``Dirac Hamiltonian in momentum space'' $\tilde{h}^{\text{Dirac}}$, as well as in the dispersion relation, Eq.\ \eqref{eq:Dirac_disp}:
\begin{equation}
k \rightarrow (- \ic) \frac{e^{\ic k a} - e^{-\ic k a} }{2 a} = \frac{\sin (k a) }{a} \, .
\end{equation}

To explain the doubling problem, it is customary to consider the following function of $k$,
\begin{empheq}[box=\widefbox]{equation}
\label{eq:naive_gapless_frequency}
f^{\text{naive}}(k) \defeq \sqrt{\big(\omega_{\pm}^{\text{naive}}(k)\big)^2 - m^2} \equiv \left| \frac{\sin (k a)}{a} \right| \, ,
\end{empheq}
to be compared with
\begin{equation}
f^{\text{Dirac}}(k)  \defeq \sqrt{\big( \omega_{\pm}^{\text{Dirac}}(k)\big)^2 - m^2} \equiv \left| k \right | \, .
\end{equation}
Notice first that the spatial discretization implies that now $k \in [-\pi/a, \pi/a )$.
Second, notice that we of course recover the continuum situation for $ka \ll 1$, because $\sin (ka) = ka + O((ka)^3)$.
We call the function $f^{\text{M}}(k)$ the gapless frequency of Model $\text{M}$.

In Fig.\ \ref{fig:the_doubling_problem}, we plot both $f^{\text{Dirac}}(k)$ and $f^{\text{naive}}(k)$ over the Brillouin zone $[-\pi/a, \pi/a )$.
The doubling problem is the following.
For a given target value $\phi_0 \defeq \sqrt{\omega_0^2 + m^2}$ of the gapless frequency, there are, in the naive discretization, not $2$ possibilities for the momentum as in the continuum situation, $k_0$ and $-k_0$  such that $f^{\text{Dirac}}(k_0) = \omega_0$, i.e., $k_0 = \omega_0$, but $4$ solutions, $2$ corresponding to the low-momentum modes that we seek to simulate with the discretization, which have $k'_0 \simeq k_0$ and $- k'_0 $ such that $f^{\text{naive}}(k'_0) = \phi_0$, and $2$ additional, high-momentum modes, namely, $\pi/a - k'_0$ and $-(\pi/a - k'_0)$, so that frequencies and momenta are not in one-to-one correspondence anymore.
In a non-interacting model, i.e., if $h$ does not depend on $x$, this is actually not a problem because the Fourier modes are independent from each other, and so the one-to-one correspondence between frequency and momentum can be tracked, e.g., fundamentally, as time evolves;
More precisely and concretely: In a non-interacting model, the momentum distribution is unchanged by the dynamics, i.e., in other words, the states of fixed momentum are stationary states.
In an interacting model, the Fourier modes will not evolve independently from each other, and the interaction term will cause the production of high-momentum modes from low-momentum ones -- because of the $2$ extra poles of $f^{\text{naive}}(k)$ with respect to $f^{\text{Dirac}}(k)$ -- which can be proved rigorously in, e.g., perturbative studies of interacting models having as zeroth order $f^{\text{naive}}(k)$ \cite{book_Rothe, Karsten1981}.

As a conclusion: In discrete space (but keeping time continuous), we will have extra, spurious modes when looking for superpositions of plane waves as Eq.\ \eqref{eq:The_solution} for the solutions.
These spurious modes are called \emph{spatial doublers}, where the specification spatial is due to the fact that what is spurious in these modes is the spatial part (high momenta, not compatible with a continuum description, even if the temporal part is -- i.e., low frequencies).

\begin{center}
\begin{figure}
\includegraphics[width = 12cm]{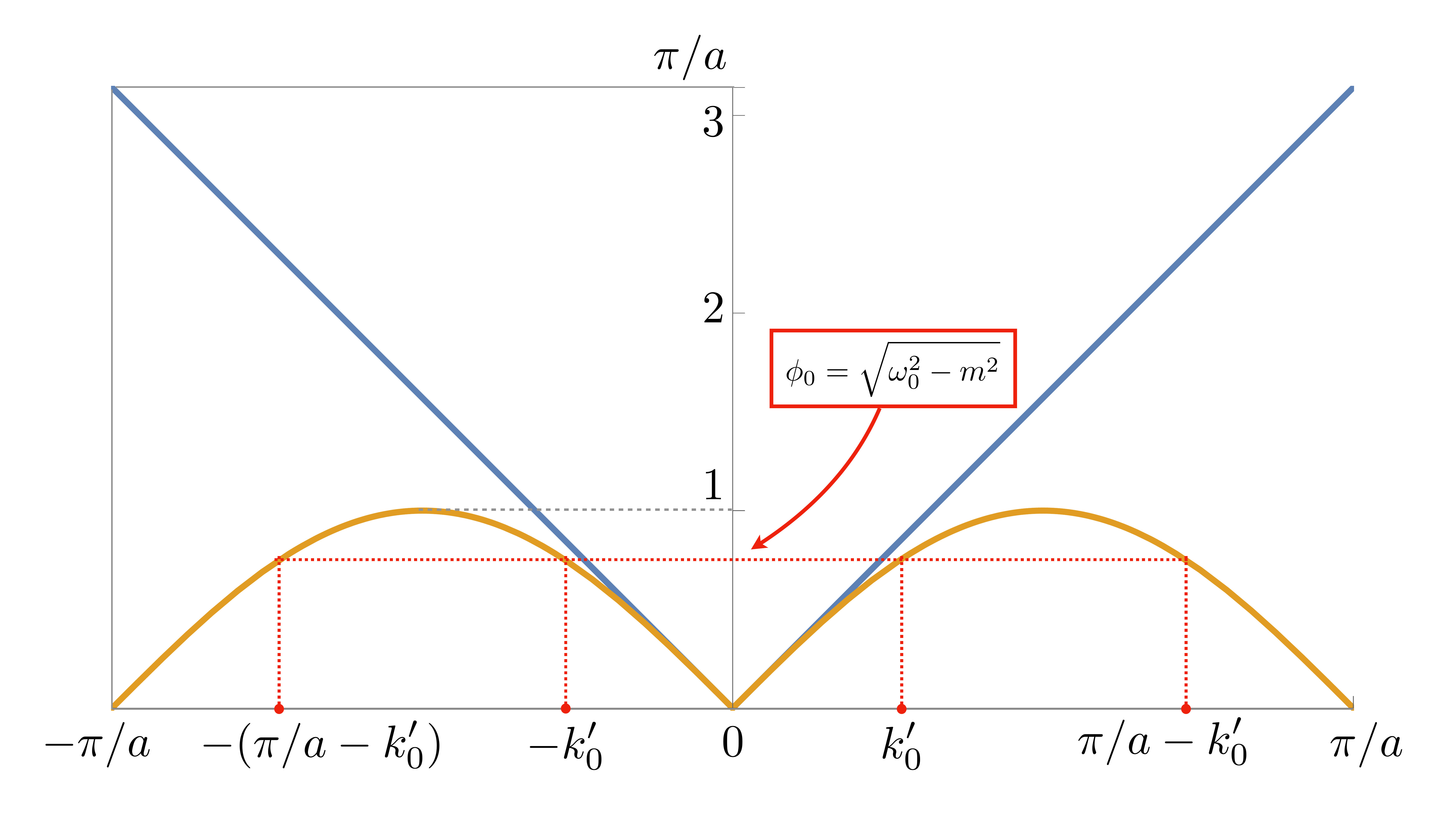}
\caption{This figures illustrates the fermion doubling problem in discrete space but continuous time. The lattice dispersion relation under the form $f_{\text{naive}}(k)$ (in gold color) has 3 poles, $-\pi/a, 0, \pi/a$, instead of a single one, $0$, for the continuum dispersion relation (in blue color), so that for each target value $\phi_0$ of the gapless frequency, we have two extra momentum solutions on the lattice with respect to the continuum situation. \label{fig:the_doubling_problem}}
\end{figure}
\end{center}

\subsection{The doubling problem in discrete spacetime: Temporal doublers in addition to the spatial doublers}
\label{subsec:temporal_doublers}

We start from the discrete-space but continuous-time situation described just above in Sec.\ \ref{subsec:continuous_time}, that is
\begin{equation}
\ic \partial_0 \Psi(\cdot,x) |_t = h^{\text{naive}} \Psi(t,\cdot) |_x \, .
\end{equation}
As we did for space in Sec.\ \ref{subsec:continuous_time}, we discretize time naively, with a symmetric finite difference, in order to treat time on the same footing as space (for which we have indeed used a symmetric finite difference), which yields (replacing $h^{\text{naive}}$ by its expression):
\begin{equation}
\label{eq:naive_scheme}
\frac{\ic}{2 \epsilon} (\Psi(t+\epsilon) - \Psi(t-\epsilon)) = \Big[ \frac{-\ic}{2 a} (e^{\ic \mathcal{K} a} - e^{-\ic \mathcal{K} a} ) \alpha^1 + m \alpha^0   \Big] \Psi(t,\cdot) |_x \, .
\end{equation}
Notice right away that this scheme takes two initial conditions, exactly as the two-step scheme we present in our paper;
The only difference with the two-step scheme we present is that in the latter there is a factor $\mu(\epsilon)$ in front of $m \alpha^0 $, but this is enough to make that scheme unitary, while the present, naive one, is not\footnote{The present scheme is indeed not unitary in the following sense. Recall first that for the unitarity of the two-step scheme of our paper has the following meaning: That two-step scheme is equivalent to the one-step scheme, for which the unitarity meaning is clear, provided that $\Psi_{j=1} = \mathcal{U} \Psi_{j=0}$. In the present case, endowing the naive two-step scheme with the same unitarity meaning does not yield a unitary one-step scheme.}.
We consider an ansatz which is a superposition of plane waves with internal components, that is, we look for solutions of the form of Eq.\ \eqref{eq:The_solution}.
If we insert Eq.\ \eqref{eq:The_solution} into Eq.\ \eqref{eq:naive_scheme}, we obtain after a few computation lines the following equation in momentum space:
\begin{equation}
\frac{\sin (\omega_i(k) \epsilon)}{\epsilon} \, \tilde{\Psi}_i(0,k) = \Big[ \frac{\sin (\omega_i(k) a)}{a} \alpha^+ m \alpha^0 \Big] \tilde{\Psi}_i(0,k) \, ,
\end{equation} 
an equation which, squared, and choosing the ballistic scaling $\epsilon = a$, finally yields the following dispersion relation,
\begin{equation}
\label{eq:temporal_doublers}
\sin^2(\omega_i(k) \epsilon) = \sin^2(k \epsilon) + \epsilon^2 m^2 \, .
\end{equation}

First of all, notice that for low frequencies $\omega_i(k) \epsilon \ll 1$ in Eq.\ \eqref{eq:temporal_doublers}, we recover the discrete-space but continuous-time situation, with gapless frequency given by Eq.\ \eqref{eq:naive_gapless_frequency} in the preceding section, Sec.\ \ref{subsec:continuous_time}.
Now, there are solutions $\omega_i(k)$ to Eq.\ \eqref{eq:temporal_doublers} if and only if $| \sin^2(k \epsilon) + \epsilon^2 m^2 | \leq 1$, which leads to
\begin{equation}
\label{eq:condition}
\epsilon^2 m^2 \leq \cos^2(k \epsilon) \, .
\end{equation}
Replacing in the dispersion relation, Eq.\ \eqref{eq:temporal_doublers}, $\sin^2(A)$ by $(1-\cos(2A))/2$, we obtain the two following solutions,
\begin{equation}
\label{eq:sol_low_freq}
\omega_{\pm}^{\text{temporal}}(k) \defeq \pm \frac{2}{\epsilon} \text{arccos}\Big( 1 - 2 \sin^2(k\epsilon) - 2 \epsilon^2 m^2 \Big) \, .
\end{equation}
For small enough $k$ and $m$, i.e., $k\epsilon \ll 1$ and $\epsilon m \ll 1$, we have that $|\omega_{\pm}^{\text{temporal}}(k) \epsilon| < \pi/2$, and actually that $|\omega_{\pm}^{\text{temporal}}(k) \epsilon| \ll 1$, and these two solutions approach the low frequencies $\omega^{\text{Dirac}}_{\pm}(k)$ of the continuum model.

Now, in addition to these two solutions, Eq.\ \eqref{eq:sol_low_freq}, we also have the solutions,
\begin{equation}
\Omega_{\pm}^{\text{temporal}}(k) \defeq \pm \Big[ \frac{\pi}{\epsilon} - \omega_{\pm}^{\text{temporal}}(k)  \Big] \, ,
\end{equation}
which are high-frequency solutions when $|\omega_{\pm}^{\text{temporal}}(k)\epsilon| < \pi/2$.
These two extra solutions are spurious because not compatible with a continuum description, but they will intervene in the dynamics in interacting models, and the modes associated to these solutions are called \emph{temporal doublers}.
As we have seen, these temporal doublers arise even for low momenta\footnote{We here speak of the momentum of the discrete model, i.e., the product $\epsilon k$.} (the only ones compatible with a continuum description), and this is best seen as follows.
Consider low momenta in Eq.\ \eqref{eq:temporal_doublers}; This yields, replacing $k$ by the notation $\kappa(w)$ and $\omega_i(k)$ by the variable $w$,
\begin{equation}
g^{\text{naive}}(w) \defeq \sqrt{\kappa(w)^2 + m^2} =  \left| \frac{\sin^2(w \epsilon)}{\epsilon} \right| \, , 
\end{equation}
an expression which, apart from the fact that there is a $+m^2$ instead of a $-m^2$, corresponds exactly to the expression of the gapless frequency, Eq.\ \eqref{eq:naive_gapless_frequency}, but having exchanged in it the roles of $\omega(k)$ and $k$, i.e., replaced the latter by $\kappa(w)$ and $w$, respectively, so that one can derive the same explanations for the temporal doublers than for the spatial doublers.
Notice that, in this low-momentum context, the condition \eqref{eq:condition} for a solution $\omega_i(k)$ to exist is
\begin{equation}
\epsilon^2 m^2 \leq 1 - \frac{1}{2} (k \epsilon)^2 \, .
\end{equation}

\subsection{Removing the doublers with Wilson fermions}
\label{subsec:continuous_time}

\subsubsection{We limit ourselves to spatial doublers}

In the preceding section, Sec.\ \ref{subsec:temporal_doublers}, we have illustrated the problem of temporal doublers starting from a continuous-time description.
A framework which is more appropriate to further remove temporal doublers is that of Lagrangian LGT, which starts from a Lagrangian continuum description rather than a Hamiltonian one.
This leads to a modification of the action rather than the Hamiltonian to solve the problem of fermion doubling, with a Wilson term, and this procedure removes the spatial doublers, as well as the the temporal ones for Wilson's choice $r=1$ \cite{Hernandez2011}.
Here, however, we will stick to a Hamiltonian formulation and treat only spatial doublers.
This is because in our paper the scheme is formulated in a Hamiltonian way.
To treat temporal doublers in the two-step scheme of our paper one would have to modify the original, DQW scheme.

\subsubsection{Treatment of spatial doublers in the naive continuous-time scheme, via Wilson fermions}

\begin{center}
\begin{figure}
\includegraphics[width=18cm]{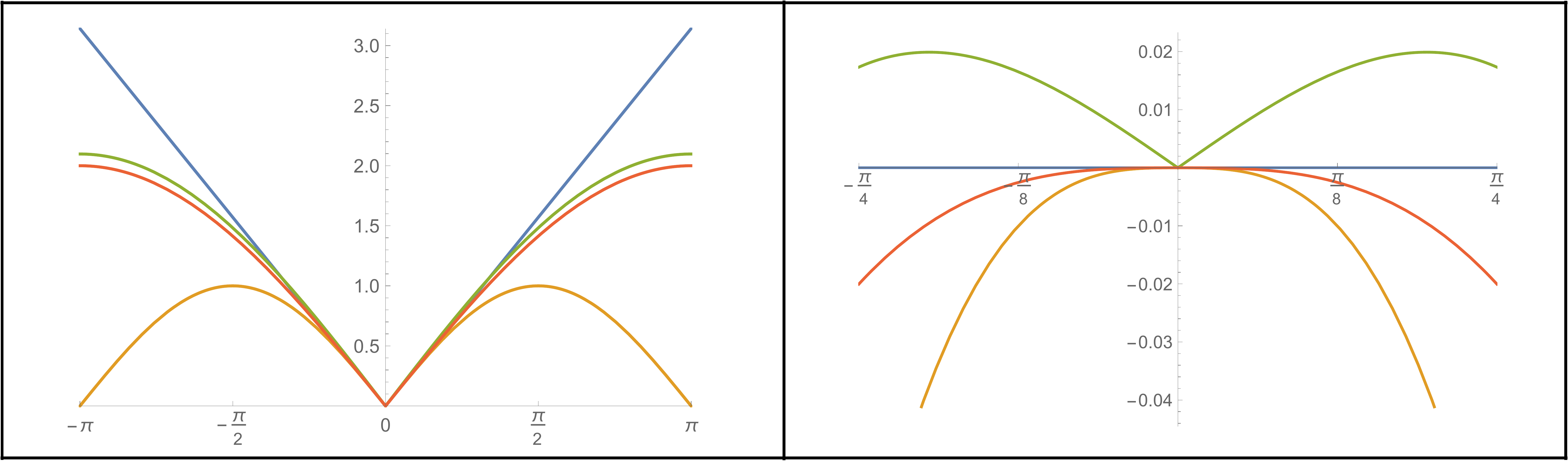}
\caption{Gapless frequency $f^{\text{M}}(k)$ of Model $\text{``M''}$ (left panel) and absolute difference $f^{\text{M}}(k) - f^{\text{Dirac}}(k)$ (right panel), for $\text{``M''} = \text{``Dirac''}$ (blue), $\text{``naive''}$ (gold), $\text{``LGT''}$ (green), and $\text{``LGT non-crossed''}$ (red), where  $f^{\text{Dirac}}(k) = |k|$, $f^{\text{naive}}(k) = \tfrac{|\sin(ka)|}{a}$,  $f^{\text{LGT}}(k) = \sqrt{\tfrac{\sin^2(ka)}{a^2} + \big[m + \tfrac{r}{a} (1 - \cos(k a)\big]^2 -m^2}$, and $f^{\text{LGT non-crossed}}(k) = \sqrt{\tfrac{\sin^2(ka)}{a^2}  + \tfrac{r^2}{a^2} \big(1 - \cos(k a) \big)^2}$, for $ a= 1$, $m=0.1$, and $r=1$ (Wilson's choice). \label{fig:doubling_fixed}}
\end{figure}
\end{center}

\begin{center}
\begin{figure}
\includegraphics[width=10cm]{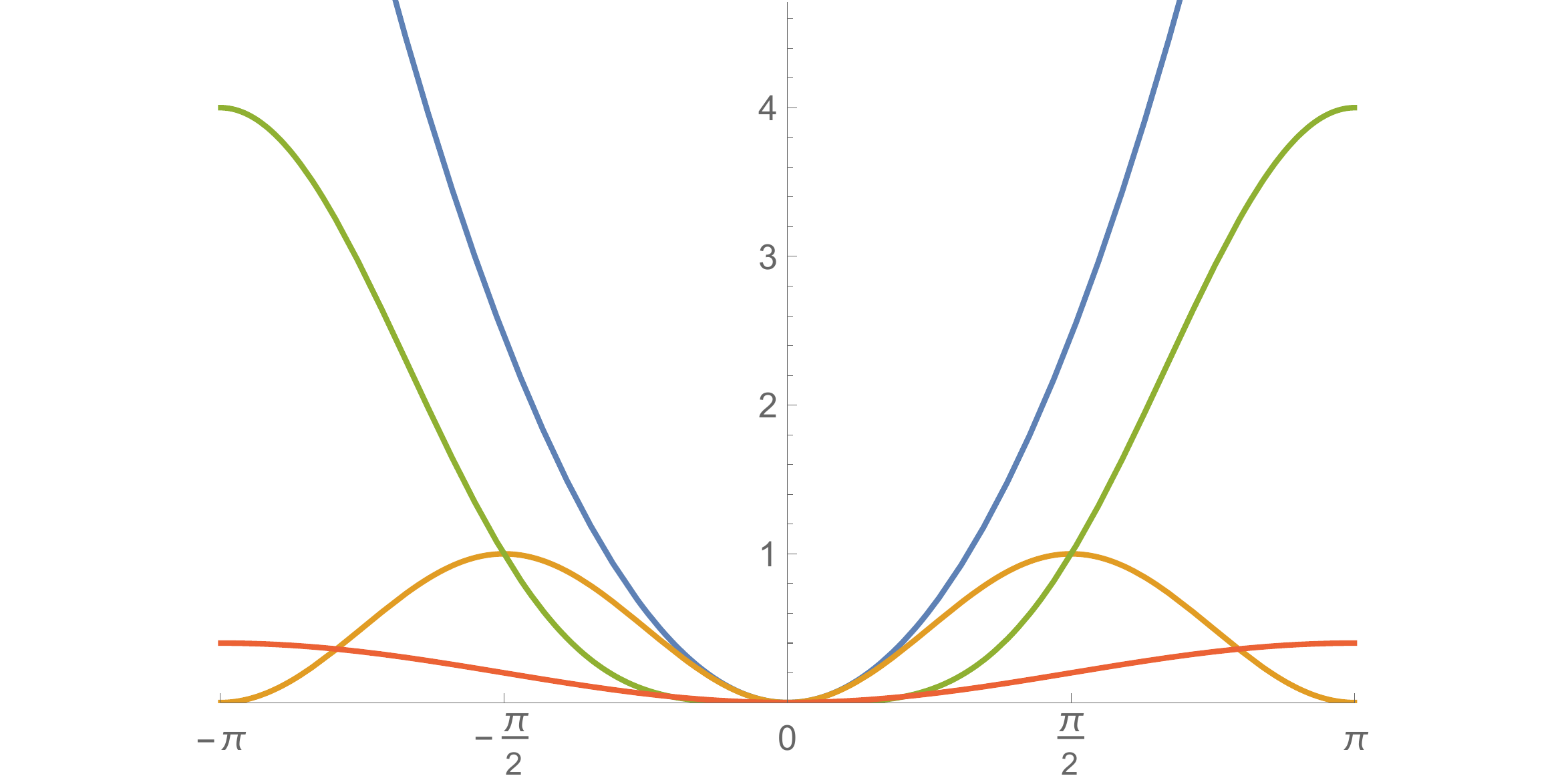}
\caption{In blue, $(f^{\text{Dirac}}(k))^2$, in gold, $(f^{\text{naive}}(k))^2$, in green, the non-crossed term $g^{\text{non-crossed}}(k)$, and in red, the crossed term $g^{\text{crossed}}(k)$. We see that the biggest contribution to the fixing of the fermion doubling comes from the non-crossed term (with respect to the crossed term). That being said, any of the two terms is actually sufficient on its own to fix the doubling problem. \label{fig:solving_the_doubling}}
\end{figure}
\end{center}

In order to treat the problem of spatial doublers in continuous-time, described in Sec.\ \ref{subsec:continuous_time}, we consider the following Hamiltonian \cite{Susskind77a},
\begin{equation}
{h}^{\text{LGT}} \defeq {h}^{\text{naive}} + {h}^{\text{Schrö.}}  \, ,
\end{equation}
where 
\begin{equation}
{h}^{\text{naive}} \defeq \alpha^1 \Big( -\ic \frac{\mathcal{D}_1}{a} \Big) + m \alpha^0 \, ,
\end{equation}
is the naive-discretization Hamiltonian considered in Sec.\ \ref{subsec:continuous_time}, and
\begin{equation}
\label{eq:Schro_term}
{h}^{\text{Schrö.}} \defeq \alpha^0 \frac{r}{a} (-\mathcal{L}) \, , 
\end{equation}
is the Wilson term, whose spatial operator, $\mathcal{L}/a$, is a discrete Laplacian\footnote{This discrete Laplacian has been obtained by replacing $a \partial_1^2$ by $\mathcal{L} \equiv d^{\ast}_1 d_1$, where $d_1$ and $d_1^{\ast}$ are respectively a forwards and a backwards finite difference (let us notice that they commute).},
\begin{equation}
\mathcal{L} = \T^{-1} + \T - 2 \, .
\end{equation}
Let us notice that the Laplacian is the spatial operator that intervenes in the non-relativistic Schrödinger equation, hence the superscript ``$\text{Schrö.}$''.

The dispersion relation is now, plotted under the form of a gapless frequency,
\begin{empheq}[box=\widefbox]{equation}
f^{\text{LGT}}(k) \defeq \sqrt{(\omega^{\text{LGT}}_{\pm})^2 - m^2} = \sqrt{\frac{\sin^2(ka)}{a^2} + \Big[ m + \frac{r}{a} \big(1 - \cos(ka) \big) \Big]^2 - m^2} \, .
\end{empheq}
In the left panel of Fig.\ \ref{fig:doubling_fixed}, we see that the doubling problem is fixed with this expression, since there are no more poles at $-\pi/a$ and $\pi/a$.
To develop explanations, it is practical to consider the square of the previous expression,
\begin{align}
\label{eq:The_square}
(f^{\text{LGT}}(k) )^2 = g^{\text{naive}}(k)   + g^{\text{non-crossed}}(k) + g^{\text{crossed}}(k) \, .
\end{align}
where
\begin{subequations}
\begin{align}
g^{\text{naive}}(k) &\defeq \frac{\sin^2(ka)}{a^2} = (f^{\text{naive}}(k))^2 \\
g^{\text{non-crossed}}(k)  &\defeq  \frac{r^2}{a^2} \big( 1 - \cos (ka) \big)^2 \\
g^{\text{crossed}}(k)  &\defeq  2 m \frac{r}{a} \big( 1 - \cos (ka) \big)  \, .
\end{align}
\end{subequations}
Let us develop this expression, Eq.\ \eqref{eq:The_square}, at next-to-next-to-lowest order in $a$,
\begin{empheq}[box=\widefbox]{equation}
\label{eq:expansion}
(f^{\text{LGT}}(k) )^2 = ( 1 + a m r )k^2 +  \Big( \frac{1}{4} a^2 r^2 -\frac{2}{3!}a^2 \Big) k^4 \, .
\end{empheq}
In Fig.\ \ref{fig:solving_the_doubling}, we see that the biggest contribution in fixing the doubling problem comes from the non-crossed term; This is actually also visible in the left panel of Fig.\ \ref{fig:doubling_fixed}. That being said, any of the two terms is actually sufficient on its own to fix the doubling problem.
Moreover, we see in Eq.\ \eqref{eq:expansion} that the crossed term unfortunately adds a first-order correction (in the lattice spacing) to the initial slope, that is, the latter becomes $1 + a m r$ instead of $1$, which is visible in the right panel of Fig.\ \ref{fig:doubling_fixed}.
Now, one can actually make this crossed term disappear, by choosing in the Wilson term $h^{\text{Schrö.}}$, Eq.\ \eqref{eq:Schro_term}, the operator $\alpha^2$ instead of $\alpha^0$.
While in continuous time this has no impact on the unitarity of the scheme, it \emph{will} a priori have an impact in discrete time \cite{Hernandez2011}, so this replacement cannot be made carelessly.

\end{widetext}

\end{document}